\documentclass[aps,prl,twocolumn,floatfix,nofootinbib,showpacs,superscriptaddress]{revtex4-1}

\usepackage{amsmath,amsfonts,amssymb,bm}
\usepackage{dsfont}
\usepackage{graphicx}
\usepackage{color}
\usepackage{soul}

\usepackage[mathscr,scaled=1.15]{urwchancal}
\DeclareFontFamily{OT1}{pzc}{}
\DeclareFontShape{OT1}{pzc}{m}{it}%
{<-> s * [1.15] pzcmi7t}{}
\DeclareMathAlphabet{\mathpzc}{OT1}{pzc}{m}{it}

\definecolor{purple}{rgb}{0.5,0,0.5}
\definecolor{blue}{rgb}{0.0,0,0.9}
\definecolor{prdblue}{rgb}{0.133,0.118,0.498}

\begin{document}

\title{Process-independent strong running coupling}
\author{Daniele Binosi}
\affiliation{European Centre for Theoretical Studies in Nuclear Physics
and Related Areas (ECT$^\ast$) and Fondazione Bruno Kessler\\ Villa Tambosi, Strada delle Tabarelle 286, I-38123 Villazzano (TN), Italy}

\author{C\'edric Mezrag}
\affiliation{Physics Division, Argonne National Laboratory, Argonne IL 60439, USA}

\author{Joannis Papavassiliou}
\affiliation{Department of Theoretical Physics and IFIC, University of Valencia and CSIC, E-46100, Valencia, Spain}

\author{Craig D. Roberts}
\affiliation{Physics Division, Argonne National Laboratory, Argonne IL 60439, USA}

\author{Jose Rodr\'{\i}guez-Quintero}
\affiliation{Department of Integrated Sciences;
University of Huelva, E-21071 Huelva, Spain}

\date{13 December 2016}

\begin{abstract}
We unify two widely different approaches to understanding the infrared behaviour of quantum chromodynamics (QCD), one essentially phenomenological, based on data, and the other computational, realised via quantum field equations in the continuum theory.  Using the latter, we explain and calculate a process-independent running-coupling for QCD, a new type of effective charge that is an analogue of the Gell-Mann--Low effective coupling in quantum electrodynamics.  The result is almost identical to the process-dependent effective charge defined via the Bjorken sum rule, which provides one of the most basic constraints on our knowledge of nucleon spin structure.  This reveals the Bjorken sum to be a near direct means by which to gain empirical insight into QCD's Gell-Mann--Low effective charge.
\end{abstract}



\maketitle

\noindent\emph{1:\;Introduction}.\,---\,In quantum gauge field theories defined in four spacetime dimensions, the Lagrangian couplings and masses do not remain constant.  Instead, owing to the need for ultraviolet (UV) renormalisation, they come to depend on a mass scale, which can often be related to the energy or momentum at which a given process occurs.  The archetype is quantum electrodynamics (QED), for which a sensible perturbation theory can be defined \cite{Nobel65}.  Within this framework, owing to the Ward identity \cite{Ward:1950xp}, there is a single running coupling, measuring the strength of the photon--charged-fermion vertex, which can be obtained by summing the collection of virtual processes that change the bare photon into a dressed object, \emph{viz}.\ by computing the photon vacuum polarisation.  QED's running coupling is known to great accuracy \cite{Olive:2016xmw} and the running has been observed directly \cite{Odaka:1998ui, Mele:2006ji}.

A new coupling appears when electromagnetism is combined with weak interactions to produce the Standard Electroweak Model \cite{Nobel79}.  It may be characterised by $\sin^2\theta_W$, where $\theta_W$ is a scale-dependent angle which specifies the particular mixing between the model's defining neutral gauge bosons that produces the observed photon and $Z^0$-boson.  A perturbation theory can also be defined for the electroweak theory \cite{Nobel99} so that $\sin^2\theta_W$ can be computed and compared with precise experiments \cite{Olive:2016xmw}.


At first sight, the addition of quantum chromodynamics (QCD) \cite{Marciano:1979wa} to the Standard Model does not qualitatively change anything, despite the presence of four possibly distinct strong-interaction vertices (gluon-ghost, three-gluon, four-gluon and gluon-quark) in the renormalised theory.  An array of Slavnov-Taylor identities (STIs) \cite{Taylor:1971ff, Slavnov:1972fg}, implementing BRST symmetry \cite{Becchi:1975nq, Tyutin:1975qk} (a generalisation of non-Abelian gauge invariance for the quantised theory) ensures that a single running coupling characterises all four interactions on the domain within which perturbation theory is valid.  The difference here is that whilst QCD is asymptotically free and extant evidence suggests that perturbation theory is valid at large momentum scales, all dynamics is nonperturbative at those scales typical of everyday strong-interaction phenomena, \emph{e.g}.\ $\zeta \lesssim m_p$, where $m_p$ is the proton's mass.

The questions that arise are how many distinct running couplings exist in nonperturbative QCD, and how can they be computed?  Given that there are four individual, apparently UV-divergent interaction vertices in the perturbative treatment of QCD, there could be as many as four distinct couplings at infrared (IR) momenta.  (Of course, if nonperturbatively there are two or more couplings, they must all become equivalent on the perturbative domain.)  In our view, nonperturbatively, too, QCD possesses a unique running coupling.  The alternative admits the possibility of a different renormalisation-group-invariant (RGI) intrinsic mass-scale for each coupling and no guarantee of a connection between them.  In such circumstances, BRST symmetry would likely be irreparably broken by nonperturbative dynamics and one would be pressed to conclude that QCD was non-renormalisable owing to IR dynamics.  There is no empirical evidence to support such a conclusion: QCD does seem to be a well-defined theory at all momentum scales, owing to the dynamical generation of gluon \cite{Cucchieri:2007md, Cucchieri:2007rg, Aguilar:2008xm, Dudal:2008sp, Bogolubsky:2009dc, Aguilar:2012rz} and quark masses \cite{Bhagwat:2003vw, Bowman:2005vx, Bhagwat:2006tu}, which are large at IR momenta.

\smallskip

\noindent\emph{2:\;Process-independent running coupling}.\,---\,Poincar\'e covariance is of enormous importance in modern physics, \emph{e.g}.\ it places severe limitations on the nature and number of those independent amplitudes that are required to fully specify any one of a gauge theory's $n$-point Schwinger functions (Euclidean Green functions).  Analyses and quantisation procedures that violate Poincar\'e covariance lead to a rapid proliferation in the number of such functions.  For example, the gluon $2$-point function (propagator, $D_{\mu\nu}$) is completely specified by one scalar function in the class of linear covariant gauges; but, in the class of axial gauges, two unconnected functions are required and unphysical, kinematic singularities are present in the associated tensors \cite{West:1982gg}.  Consequently, covariant gauges are typically preferred for concrete calculations in both continuum and lattice-regularised studies of QCD.  In fact, Landau gauge is the most common choice because, \emph{inter alia}, it is a fixed point of the renormalisation group and readily implemented in lattice-QCD \cite{Cucchieri:2009kk}.  Herein, therefore, we use Landau gauge; and, moreover, employ a physical momentum-subtraction renormalisation scheme, detailed elsewhere \cite{Aguilar:2009nf}.

As noted in Sec.\,1, there is a particular simplicity to QED, \emph{viz}.\ the unique running coupling, a process-independent effective charge, can be obtained simply by computing the photon vacuum polarisation.  This is because ghost-fields decouple in Abelian theories; and, consequently, one has the Ward identity, which guarantees that the electric-charge renormalisation constant is equivalent to that of the photon field.  Stated physically, the impact of dressing the interaction vertices is absorbed into the vacuum polarisation.  This is not generally true in QCD because ghost-fields do not decouple.

There is one approach to analysing QCD's Schwinger functions, however, that preserves some of QED's simplicity; namely, the combination of pinch technique (PT) \cite{Cornwall:1981zr, Cornwall:1989gv, Pilaftsis:1996fh, Binosi:2002ft, Binosi:2003rr, Binosi:2009qm} and background field method (BFM) \cite{Abbott:1980hw, Abbott:1981ke}.   This framework can be seen as a means by which QCD can be made to ``look'' Abelian: one systematically rearranges classes of diagrams and their sums in order to obtain modified Schwinger functions that satisfy linear STIs.  In the gauge sector, in Landau gauge, this produces a modified gluon dressing function from which one can compute the QCD running coupling, \emph{i.e}.\ the polarisation captures all required features of the renormalisation group.  Furthermore, the coupling is process independent: one obtains precisely the same result, independent of the scattering process considered, whether gluon+gluon$\,\to\,$gluon+gluon, quark+quark$\,\to\,$quark+quark, etc.  This clean connection between the coupling and the gluon vacuum polarisation relies on another particular feature of QCD, \emph{viz}.\ in Landau gauge the renormalisation constant of the gluon-ghost vertex is not only finite but unity \cite{Taylor:1971ff}, in consequence of which the effective charge obtained from the PT-BFM gluon vacuum polarisation is directly connected with that deduced from the gluon-ghost vertex \cite{Aguilar:2009nf}, sometimes called the ``Taylor coupling,'' $\alpha_{\rm T}$ \cite{Blossier:2011tf, Blossier:2012ef, Blossier:2013ioa}.

Writing these statements explicitly, with $T_{\mu\nu}(k)=\delta_{\mu\nu}-k_\mu k_\nu/k^2$, one has \cite{Binosi:2014aea, Binosi:2016xxu}
\begin{subequations}
\label{allhatd}
\begin{align}
\label{hatd}
\alpha(\zeta^2) D^{\rm PB}_{\mu\nu}(k;\zeta) & = \widehat{d}(k^2)\, T_{\mu\nu}(k)\,,\\
%
%
  \mathpzc{I}(k^2) :=k^2 \widehat{d}(k^2) &= \frac{\alpha_{\rm T}(k^2)}{[1-L(k^2;\zeta^2)F(k^2;\zeta^2)]^2}\,,
  \label{mathcalI}
\end{align}
\end{subequations}
where:
$\alpha(\zeta^2)=g^2(\zeta^2)/[4\pi]$, $\zeta$ is the renormalisation scale;
$D^{\rm PB}_{\mu\nu}$ is the PT-BFM gluon two-point function;
$\widehat{d}(k^2)$ is the RGI running-interaction discussed in Ref.\,\cite{Aguilar:2009nf};
$F$ is the dressing function for the ghost propagator;
and $L$ is a longitudinal piece of the gluon-ghost vacuum polarisation that vanishes at $k^2=0$.
In terms of these quantities, QCD's matter-sector gap equation can be written $(k=p-q)$
\begin{subequations}
\label{gendseN}
\begin{align}
S^{-1}(p) 
& = Z_2 \,(i\gamma\cdot p + m^{\rm bm}) + \Sigma(p)\,,\\
\Sigma(p)& =  Z_2\int^\Lambda_{dq}\!\!
4\pi \widehat{d}(k^2) \,T_{\mu\nu}(k)\gamma_\mu S(q) \hat\Gamma^a_\nu(q,p)\, ,
\end{align}
\end{subequations}
where the usual $Z_1 \Gamma^a_\nu$ has become  $Z_2 \hat\Gamma^a_\nu$, with the latter being a PT-BFM gluon-quark vertex that satisfies an Abelian-like Ward-Green-Takahashi identity \cite{Binosi:2009qm} and $Z_{1,2}$ are, respectively, the gluon-quark vertex and quark wave function renormalisation constants.

The RGI interaction, $\widehat{d}(k^2)$, in Eqs.\,\eqref{allhatd} has been computed.  The most up-to-date result is discussed in Refs.\,\cite{Binosi:2014aea, Binosi:2016xxu}.  These analyses make explicit a remarkable feature of QCD; namely, the interaction saturates at infrared momenta:
\begin{equation}
\widehat{d}(k^2=0) = \alpha(\zeta^2)/m_g^2(\zeta)=\alpha_0/m_0^2\,,
\end{equation}
where $\alpha_0:=\alpha(0) \approx 0.9\pi$, $m_0:=m_g(0) \approx m_p/2$, \emph{i.e}.\ the gluon sector of QCD is characterised by a nonperturbatively-generated infrared mass-scale \cite{Cucchieri:2007md, Cucchieri:2007rg, Aguilar:2008xm, Dudal:2008sp, Bogolubsky:2009dc, Aguilar:2012rz}.  With this in mind, we define a RGI function
\begin{align}
\label{calD}
\mathpzc{D}(k^2) & = \Delta_{\rm F}(k^2;\zeta) \, m_g^2(\zeta^2)/m_0^2 \,,
\end{align}
where $\Delta_{\rm F}$ is a parametrisation of continuum- and/or lattice-QCD calculations of the canonical gluon two-point function, built such that the IR behaviour is preserved and $1/\Delta_{\rm F}(k^2;\zeta) = k^2+{\rm O}(1)$ on $k^2\gg m_0^2$.  Using Eq.\,\eqref{calD},
\begin{equation}
\Sigma(p)  =   Z_2 \int^\Lambda_{dq}\!\!
4\pi \widehat{\alpha}_{\rm PI}(k^2) \mathpzc{D}_{\mu\nu}(k^2) \gamma_\mu S(q) \hat\Gamma^a_\nu(q,p)\,,
\end{equation}
where $\mathpzc{D}_{\mu\nu} =\mathpzc{D} T_{\mu\nu}$ and the dimensionless product
\begin{equation}
\widehat{\alpha}_{\rm PI}(k^2)  = \widehat{d}(k^2) / \mathpzc{D}(k^2)
\label{widehatalpha}
\end{equation}
is a RGI running-coupling (effective charge): by construction, $\widehat{\alpha}_{\rm PI}(k^2) = \mathpzc{I}(k^2)$ on $k^2\gg m_0^2$.

The product in Eq.\,\eqref{widehatalpha} has many important qualities.  For instance, it is process independent: as noted above, the same function appears irrespective of the initial and final parton systems.
Moreover, it unifies a diverse and extensive array of hadron observables \cite{Binosi:2014aea}; a property that is evident in the fact that the dressed-quark self-energy serves as a generating functional for the Bethe-Salpeter kernel in all meson channels and the product $\widehat{\alpha}_{\rm PI}(k^2)$ is untouched by the generating procedure in all flavoured systems \cite{Munczek:1994zz, Bender:1996bb, Bhagwat:2007ha, Binosi:2016rxz}.
Finally, although $\widehat{\alpha}_{\rm PI}(k^2)$ is RGI and process-independent in any gauge, it is sufficient to know $\widehat{\alpha}_{\rm PI}(k^2)$ in Landau gauge (the choice for easiest computation) because $\widehat{\alpha}_{\rm PI}(k^2)$ is form-invariant under gauge transformations \cite{Binosi:2013cea} and, crucially, gauge covariance ensures that such transformations produce nothing but an overall ``phase'' in the gap equation's solution, which may be absorbed into the dressed-quark two-point function.

\smallskip

\begin{figure}[t]

\centerline{
\includegraphics[clip,width=0.44\textwidth]{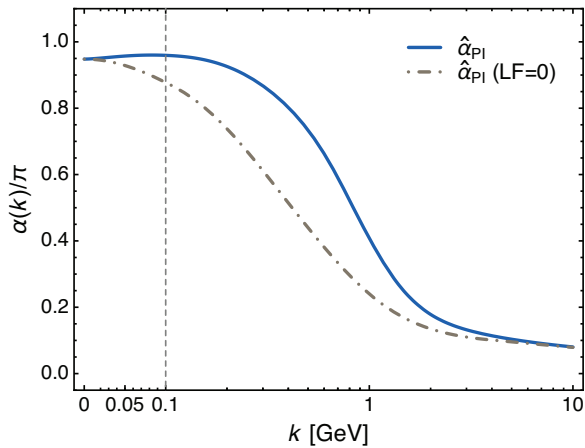}}

\caption{\label{Figwidehatalpha}
Solid (blue) curve, complete effective charge in Eq.\,\eqref{widehatalpha}; and dot-dashed (black) curve, Taylor-scheme effective charge, \emph{i.e}.\ computed in the absence of crucial pieces of the gluon-ghost vacuum polarisation [$L F \equiv 0$ in Eq.\,\eqref{mathcalI}].
The k-axis scale is linear to the left of the vertical line and logarithmic otherwise, an artifice which enables us to show saturation of the effective charge.}
\end{figure}

\noindent\emph{3:\;Computing the running coupling}.\,---\,The effective charge defined in Eq.\,\eqref{widehatalpha} is a product of known quantities: both $\widehat{d}(k^2)$  and the canonical gluon two-point function have been extensively studied and tightly constrained using continuum and lattice methods \cite{Binosi:2014aea, Binosi:2016wcx, Binosi:2016xxu}.  Indeed, the known forms of these functions provide a unified, quantitatively reliable explanation of numerous hadron physics observables \cite{Binosi:2014aea, Binosi:2016wcx}.  It is therefore straightforward to combine existing results and compute $\widehat{d}(k^2)$, a procedure \cite{Binosi:2016xxu} which yields the function depicted in Fig.\,\ref{Figwidehatalpha}.  For this purpose we used a $[n,n+1]$, $n=1$, Pad\'e approximant to simultaneously interpolate the IR behaviour of contemporary lattice results for $D_{\mu\nu}(k)$ \cite{Binosi:2016xxu} and express the UV constraint on $\Delta_{\rm F}(k^2;\zeta)$.  (Using $n\geq 2$ yields no noticeable improvement, but $n=0$ is incapable of representing modern lattice data.)

It is worth highlighting some important features of the effective charge in Fig.\,\ref{Figwidehatalpha}.
First, it is a parameter-free prediction: the curve is completely determined by results obtained for the gluon and ghost two-point functions using continuum and lattice-regularised QCD.
Second, it is physical, in the sense that there is no Landau pole, and it saturates in the IR: $\widehat{\alpha}_{\rm PI}(k^2=0) =\alpha_0 \approx 0.9 \pi$, \emph{i.e}.\ the coupling possesses an infrared fixed point \cite{Aguilar:2002tc}.
Third, the prediction is equally concrete and sound at all spacelike momenta, connecting the IR and UV domains, and precisely reproducing the known behaviour of the Taylor coupling at large $k^2$ \cite{Blossier:2011tf, Blossier:2012ef, Blossier:2013ioa}, with no need for an \emph{ad hoc} ``matching procedure,'' such as that employed in models \cite{Deur:2016tte}.
Finally, our result is essentially nonperturbative, obtained by combining self-consistent solutions of gauge-sector gap equations with lattice simulations, augmented only by a physical procedure for setting a single mass-scale \cite{Binosi:2016xxu}.
There are indications \cite{Aguilar:2001zy, Natale:2009uz, Luna:2010tp} that the effective charge in Fig.\,\ref{Figwidehatalpha} could prove useful in developing a modern dynamical perturbation theory \cite{Pagels:1979hd}.

It is evident in Fig.\,\ref{Figwidehatalpha} that ghost-gluon interactions are critical.  The RGI product $L F$ in Eq.\,\eqref{mathcalI} expresses effects of gluon-ghost scattering that are essential to ensuring $\widehat{\alpha}_{\rm PI}$ is process-independent.  It is also quantitatively important, introducing a roughly 60\% enhancement of $\widehat{\alpha}_{\rm PI}(k^2) $ for $k\simeq m_0$.  It must also, therefore, be physically significant because the strength of the running coupling at IR momenta determines the magnitude of dynamical chiral symmetry breaking (DCSB)  \cite{Binosi:2014aea, Binosi:2016wcx, Binosi:2016xxu}; and DCSB is a crucial emergent phenomenon in QCD, possibly inseparable from confinement in the unquenched theory \cite{Horn:2016rip}, \emph{i.e}. when dynamical light quarks are active.

\smallskip

\noindent\emph{4:\;Comparison of effective charges}.\,---\,Another approach to determining an ``effective charge'' in QCD was introduced in Ref.\,\cite{Grunberg:1982fw}.  This is a process-dependent procedure; namely, an effective running coupling is defined to be completely fixed by the leading-order term in the perturbative expansion of a given observable in terms of the canonical running coupling.  An obvious difficulty, or perhaps drawback, of such a scheme is the process-dependence itself.  Naturally, effective charges from different observables can in principle be algebraically connected to each other via an expansion of one coupling in terms of the other.  However, any such expansion contains infinitely many terms \cite{Deur:2016tte}; and this connection does not imbue a given process-dependent charge with the ability to predict any other observable, since the expansion is only defined \emph{a posteriori}, \emph{i.e}.\ after both effective charges are independently constructed.

One such process-dependent effective charge is $\alpha_{g_1}(k^2)$, which is defined via the Bjorken sum rule \cite{Bjorken:1966jh, Bjorken:1969mm}:
\begin{align}
\int_0^1 \! dx  \left[g_1^p(x,k^2) - g_1^n(x,k^2)\right] = \frac{g_A}{6}  \left[ 1 - \tfrac{1}{\pi} \alpha_{g_1}(k^2) \right]\,,
\end{align}
where $g_1^{p,n}$ are the spin-dependent proton and neutron structure functions, whose extraction requires measurements using polarised targets, 
and $g_A$ is the nucleon flavour-singlet axial-charge \cite{Aidala:2012mv}.  The merits of this definition are outlined in Ref.\,\cite{Deur:2016tte}.  They include the existence of data for a wide range of $k^2$ \cite{%
Deur:2005cf, Deur:2008rf, Deur:2014vea,
Ackerstaff:1997ws, Ackerstaff:1998ja, Airapetian:1998wi, Airapetian:2002rw, Airapetian:2006vy,
Kim:1998kia,
Alexakhin:2006oza, Alekseev:2010hc, Adolph:2015saz,
Anthony:1993uf, Abe:1994cp, Abe:1995mt, Abe:1995dc, Abe:1995rn, Anthony:1996mw, Abe:1997cx, Abe:1997qk, Abe:1997dp, Abe:1998wq, Anthony:1999py, Anthony:1999rm, Anthony:2000fn, Anthony:2002hy};
tight sum-rules constraints on the behaviour of the integral at the IR and UV extremes of $k^2$;
and the isospin non-singlet feature of the difference, which suppresses contributions from numerous processes that are hard to compute and hence might muddy interpretation of the integral in terms of an effective charge.


\begin{figure}[t]

\centerline{
\includegraphics[clip,width=0.48\textwidth]{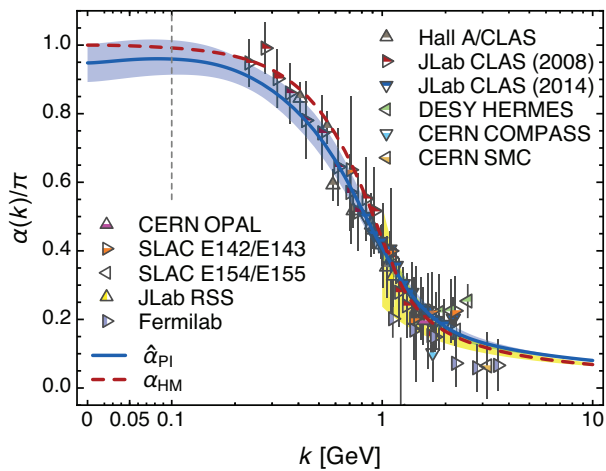}}

\caption{\label{FigwidehatalphaII}
Solid (blue) curve: predicted process-in\-de\-pen\-dent RGI running-coupling $\widehat{\alpha}_{\rm PI}(k^2)$, Eq.\,\eqref{widehatalpha}.  The shaded (blue) band bracketing this curve combines a 95\% confidence-level window based on existing lattice-QCD results for the gluon two-point function with an error of 10\% in the continuum extraction of the RGI product $L F$ in Eqs.\,\eqref{allhatd}.
World data on $\alpha_{g_1}$ \cite{%
Deur:2005cf, Deur:2008rf, Deur:2014vea,
Ackerstaff:1997ws, Ackerstaff:1998ja, Airapetian:1998wi, Airapetian:2002rw, Airapetian:2006vy,
Kim:1998kia,
Alexakhin:2006oza, Alekseev:2010hc, Adolph:2015saz,
Anthony:1993uf, Abe:1994cp, Abe:1995mt, Abe:1995dc, Abe:1995rn, Anthony:1996mw, Abe:1997cx, Abe:1997qk, Abe:1997dp, Abe:1998wq, Anthony:1999py, Anthony:1999rm, Anthony:2000fn, Anthony:2002hy}.
The shaded (yellow) band on $k>1\,$GeV represents $\alpha_{g_1}$ obtained from the Bjorken sum by using QCD evolution \cite{Gribov:1972, Altarelli:1977, Dokshitzer:1977} to extrapolate high-$k^2$ data into the depicted region, following Refs.\,\cite{Deur:2005cf, Deur:2008rf}; and, for additional context, the dashed (red) curve is the light-front holographic model of $\alpha_{g_1}$ canvassed in Ref.\,\cite{Deur:2016tte}.
%
}
\end{figure}

The world's data on the process-dependent effective charge $\alpha_{g_1}(k^2)$ are depicted in Fig.\,\ref{FigwidehatalphaII} and therein compared with our prediction for the process-independent RGI running-coupling $\widehat{\alpha}_{\rm PI}(k^2)$.
Owing to asymptotic freedom, all reasonable definitions of a QCD effective charge must agree on $k^2\gtrsim 1\,$GeV$^2$ and our approach guarantees this connection.  To be specific, in terms of the widely-used $\overline{\rm MS}$ running coupling \cite{Olive:2016xmw}:
\begin{subequations}
\label{AgreeCouplings}
 \begin{align}
 \label{ag1MSbar}
 \alpha_{g_1}(k^2) & = \alpha_{\overline{\rm MS}}(k^2) ( 1 + 1.14 \, \alpha_{\overline{\rm MS}}(k^2) + \ldots ) \,,\\
 \widehat{\alpha}_{\rm PI}(k^2) & = \alpha_{\overline{\rm MS}}(k^2) ( 1 + 1.09 \, \alpha_{\overline{\rm MS}}(k^2) + \ldots ) \,,
 \end{align}
\end{subequations}
where Eq.\,\eqref{ag1MSbar} may be built from, \emph{e.g}.\ Refs.\,\cite{Kataev:1994gd, Baikov:2010je}.

Significantly, there is also near precise agreement with data on the IR domain, $k^2 \lesssim m_0^2$, and complete accord on $k^2 \geq m_0^2$.  Fig.\,\ref{Figwidehatalpha} makes plain that any agreement on $k^2\in[0.01,1]\,$GeV$^2$ is non-trivial because ghost-gluon interactions produce as much as 40\% of $\widehat{\alpha}_{\rm PI}(k^2)$ on this domain: if these effects were omitted from the gluon vacuum polarisation, then $\alpha_{g_1}$ and $\widehat{\alpha}_{\rm PI}$ would differ by roughly a factor of two on the critical domain of transition between strong and perturbative QCD.


\smallskip

\noindent\emph{5:\;Conclusions}.\,---\,We have defined and calculated a process-independent running-coupling for QCD, $\widehat{\alpha}_{\rm PI}(k^2)$ [Eq.\,\eqref{widehatalpha}, Fig.\,\ref{Figwidehatalpha}].  This is a new type of effective charge, which is an analogue of the Gell-Mann--Low effective coupling in QED, being completely determined by the gauge-boson two-point function.
Our prediction for $\widehat{\alpha}_{\rm PI}(k^2)$ is parameter-free, being obtained by combining the self-consistent solution of a set of Dyson-Schwinger equations with results from lattice-QCD; and it smoothly unifies the nonperturbative and perturbative domains of the strong-interaction theory. 
This process-independent running coupling is known to unify a vast array of observables, \emph{e.g}.\ the pion mass and decay constant, and the light meson spectrum \cite{Chang:2011ei}; the parton distribution amplitudes of light- and heavy-mesons \cite{Chang:2013pq, Shi:2015esa, Ding:2015rkn}, associated elastic and transition form factors \cite{Raya:2015gva, Raya:2016yuj}, etc.

Finally, and perhaps surprisingly at first sight, $\widehat{\alpha}_{\rm PI}(k^2)$ is almost pointwise identical at infrared momenta to the process-dependent effective charge, $\alpha_{g_1}$, defined via the Bjorken sum rule, one of the most basic constraints on our knowledge of nucleon spin structure, and in complete agreement on the domain of perturbative momenta [Fig.\,\ref{FigwidehatalphaII}].  Equivalence on the perturbative domain is guaranteed for any two reasonable definitions of QCD's effective charge, but here the subleading terms differ by just 4\% [Eqs.\,\eqref{AgreeCouplings}].  An excellent match at infrared momenta, \emph{i.e}.\ below the scale at which perturbation theory would locate the Landau pole, is non-trivial; and crucial to this agreement is the careful treatment and incorporation of a special class of gluon-ghost scattering effects.
One is naturally compelled to ask how these two apparently unrelated definitions of a QCD effective charge can be so similar?  We attribute this outcome to a physically useful feature of the Bjorken sum rule, \emph{viz}.\ it is an isospin non-singlet relation and hence contributions from many hard-to-compute processes are suppressed, and these same processes are omitted in our computation of $\widehat{\alpha}_{\rm PI}(k^2)$.

The analysis herein unifies two vastly different approaches to understanding the infrared behaviour of QCD, one essentially phenomenological and the other deliberately computational, embedded within QCD.  There is no Landau pole in our predicted running coupling.  In fact, there is an inflection point at $\surd k^2=0.7\,$GeV, marking a transition wall at which, as momenta decreasing from the ultraviolet promote growth in the coupling, that coupling turns away from the Landau pole, the growth slows, and finally the coupling saturates: $\widehat\alpha_{\rm PI}(k^2=0) \approx 0.9\pi$  [Fig.\,\ref{FigwidehatalphaII}].  This unification identifies the Bjorken sum rule as a near direct means by which to gain empirical insight into a QCD analogue of the Gell-Mann--Low effective charge.

\smallskip

%
\noindent\emph{Acknowledgments}.\,---\,We are grateful for comments from S.\,J.\,Brodsky, L.\,Chang, A.\,Deur and S.-X.\,Qin.
%
This study was conceived and initiated during the $3^{rd}$ Workshop on Non-perturbative QCD, University of Seville, Spain, 17-21 October 2016.
This research was supported by:
Spanish MEYC, under grants FPA2014-53631-C-1-P, FPA2014-53631-C-2-P and SEV-2014-0398;
Generalitat Valenciana  under grant Prometeo~II/2014/066;
and
U.S.\ Department of Energy, Office of Science, Office of Nuclear Physics, contract no.~DE-AC02-06CH11357.
%


\begin{thebibliography}{10}

\bibitem{Nobel65}
S.~Lundqvist~(Editor),
\newblock {\em Nobel Lectures in Physics (1963-1970)} (World Scientific,
  Singapore, 1998), pp. 121--180.

\bibitem{Ward:1950xp}
J.~C. Ward,
\newblock Phys. Rev. {\bf 78}, 182 (1950).

\bibitem{Olive:2016xmw}
C.~Patrignani {\em et~al.},
\newblock Chin. Phys. C {\bf 40}, 100001 (2016).

\bibitem{Odaka:1998ui}
S.~Odaka {\em et~al.},
\newblock Phys. Rev. Lett. {\bf 81}, 2428 (1998).

\bibitem{Mele:2006ji}
S.~Mele,
\newblock (hep-ex/0610037),
\newblock {Measurements of the running of the electromagnetic coupling at LEP}.

\bibitem{Nobel79}
S.~Lundqvist~(Editor),
\newblock {\em Nobel Lectures in Physics (1971-1980)} (World Scientific,
  Singapore, 1994), pp. 485--560.

\bibitem{Nobel99}
G.~Ekspong~(Editor),
\newblock {\em Nobel Lectures in Physics (1996-2000)} (World Scientific,
  Singapore, 2002), pp. 359--397.

\bibitem{Marciano:1979wa}
W.~J. Marciano and H.~Pagels,
\newblock Nature {\bf 279}, 479 (1979).

\bibitem{Taylor:1971ff}
J.~C. Taylor,
\newblock Nucl. Phys. B {\bf 33}, 436 (1971).

\bibitem{Slavnov:1972fg}
A.~A. Slavnov,
\newblock Theor. Math. Phys. {\bf 10}, 99 (1972).

\bibitem{Becchi:1975nq}
C.~Becchi, A.~Rouet and R.~Stora,
\newblock Annals Phys. {\bf 98}, 287 (1976).

\bibitem{Tyutin:1975qk}
I.~V. Tyutin,
\newblock (1975),
\newblock {Gauge Invariance in Field Theory and Statistical Physics in Operator
  Formalism, arXiv:0812.0580 [hep-th]}.

\bibitem{Cucchieri:2007md}
A.~Cucchieri and T.~Mendes,
\newblock PoS {\bf LAT2007}, 297 (2007).

\bibitem{Cucchieri:2007rg}
A.~Cucchieri and T.~Mendes,
\newblock Phys. Rev. Lett. {\bf 100}, 241601 (2008).

\bibitem{Aguilar:2008xm}
A.~Aguilar, D.~Binosi and J.~Papavassiliou,
\newblock Phys. Rev. D {\bf 78}, 025010 (2008).

\bibitem{Dudal:2008sp}
D.~Dudal, J.~A. Gracey, S.~P. Sorella, N.~Vandersickel and H.~Verschelde,
\newblock Phys. Rev. {\bf D78}, 065047 (2008).

\bibitem{Bogolubsky:2009dc}
I.~Bogolubsky, E.~Ilgenfritz, M.~Muller-Preussker and A.~Sternbeck,
\newblock Phys. Lett. B {\bf 676}, 69 (2009).

\bibitem{Aguilar:2012rz}
A.~Aguilar, D.~Binosi and J.~Papavassiliou,
\newblock Phys. Rev. D {\bf 86}, 014032 (2012).

\bibitem{Bhagwat:2003vw}
M.~S. Bhagwat, M.~A. Pichowsky, C.~D. Roberts and P.~C. Tandy,
\newblock Phys. Rev. C {\bf 68}, 015203 (2003).

\bibitem{Bowman:2005vx}
P.~O. Bowman {\em et~al.},
\newblock Phys. Rev. D {\bf 71}, 054507 (2005).

\bibitem{Bhagwat:2006tu}
M.~S. Bhagwat and P.~C. Tandy,
\newblock AIP Conf. Proc. {\bf 842}, 225 (2006).

\bibitem{West:1982gg}
G.~B. West,
\newblock Phys. Rev. D {\bf 27}, 1878 (1983).

\bibitem{Cucchieri:2009kk}
A.~Cucchieri, T.~Mendes and E.~M.~S. Santos,
\newblock Phys. Rev. Lett. {\bf 103}, 141602 (2009).

\bibitem{Aguilar:2009nf}
A.~Aguilar, D.~Binosi, J.~Papavassiliou and J.~Rodr{\'i}guez-Quintero,
\newblock Phys. Rev. D {\bf 80}, 085018 (2009).

\bibitem{Cornwall:1981zr}
J.~M. Cornwall,
\newblock Phys. Rev. D {\bf 26}, 1453 (1982).

\bibitem{Cornwall:1989gv}
J.~M. Cornwall and J.~Papavassiliou,
\newblock Phys. Rev. D {\bf 40}, 3474 (1989).

\bibitem{Pilaftsis:1996fh}
A.~Pilaftsis,
\newblock Nucl. Phys. B {\bf 487}, 467 (1997).

\bibitem{Binosi:2002ft}
D.~Binosi and J.~Papavassiliou,
\newblock Phys. Rev. D {\bf 66}, 111901 (2002).

\bibitem{Binosi:2003rr}
D.~Binosi and J.~Papavassiliou,
\newblock J. Phys. G {\bf 30}, 203 (2004).

\bibitem{Binosi:2009qm}
D.~Binosi and J.~Papavassiliou,
\newblock Phys. Rept. {\bf 479}, 1 (2009).

\bibitem{Abbott:1980hw}
L.~F. Abbott,
\newblock Nucl. Phys. B {\bf 185}, 189 (1981).

\bibitem{Abbott:1981ke}
L.~F. Abbott,
\newblock Acta Phys. Polon. B {\bf 13}, 33 (1982).

\bibitem{Blossier:2011tf}
B.~Blossier {\em et~al.},
\newblock Phys. Rev. D {\bf 85}, 034503 (2012).

\bibitem{Blossier:2012ef}
B.~Blossier {\em et~al.},
\newblock Phys. Rev. Lett. {\bf 108}, 262002 (2012).

\bibitem{Blossier:2013ioa}
B.~Blossier {\em et~al.},
\newblock Phys. Rev. D {\bf 89}, 014507 (2014).

\bibitem{Binosi:2014aea}
D.~Binosi, L.~Chang, J.~Papavassiliou and C.~D. Roberts,
\newblock Phys. Lett. B {\bf 742}, 183 (2015).

\bibitem{Binosi:2016xxu}
D.~Binosi, C.~D. Roberts and J.~Rodriguez-Quintero,
\newblock (arXiv:1611.03523 [nucl-th]),
\newblock {\emph{Scale-setting, flavour dependence and chiral symmetry
  restoration}}.

\bibitem{Munczek:1994zz}
H.~J. Munczek,
\newblock Phys. Rev. D {\bf 52}, 4736 (1995).

\bibitem{Bender:1996bb}
A.~Bender, C.~D. Roberts and L.~von Smekal,
\newblock Phys. Lett. B {\bf 380}, 7 (1996).

\bibitem{Bhagwat:2007ha}
M.~S. Bhagwat, L.~Chang, Y.-X. Liu, C.~D. Roberts and P.~C. Tandy,
\newblock Phys. Rev. C {\bf 76}, 045203 (2007).

\bibitem{Binosi:2016rxz}
D.~Binosi, L.~Chang, J.~Papavassiliou, S.-X. Qin and C.~D. Roberts,
\newblock Phys. Rev. D {\bf 93}, 096010 (2016).

\bibitem{Binosi:2013cea}
D.~Binosi and A.~Quadri,
\newblock Phys. Rev. D {\bf 88}, 085036 (2013).

\bibitem{Binosi:2016wcx}
D.~Binosi, L.~Chang, J.~Papavassiliou, S.-X. Qin and C.~D. Roberts,
\newblock (arXiv:1609.02568 [nucl-th]),
\newblock {\emph{Natural constraints on the gluon-quark vertex}}.

\bibitem{Aguilar:2002tc}
A.~C. Aguilar, A.~A. Natale and P.~S. Rodrigues~da Silva,
\newblock Phys. Rev. Lett. {\bf 90}, 152001 (2003).

\bibitem{Deur:2016tte}
A.~Deur, S.~J. Brodsky and G.~F. de~Teramond,
\newblock Prog. Part. Nucl. Phys. {\bf 90}, 1 (2016).

\bibitem{Aguilar:2001zy}
A.~C. Aguilar, A.~Mihara and A.~A. Natale,
\newblock Phys. Rev. {\bf D65}, 054011 (2002).

\bibitem{Natale:2009uz}
A.~A. Natale,
\newblock PoS {\bf QCD-TNT09}, 031 (2009).

\bibitem{Luna:2010tp}
E.~G.~S. Luna, A.~L. dos Santos and A.~A. Natale,
\newblock Phys. Lett. B {\bf 698}, 52 (2011).

\bibitem{Pagels:1979hd}
H.~Pagels and S.~Stokar,
\newblock Phys. Rev. D {\bf 20}, 2947 (1979).

\bibitem{Horn:2016rip}
T.~Horn and C.~D. Roberts,
\newblock J. Phys. G. {\bf 43}, 073001/1 (2016).

\bibitem{Grunberg:1982fw}
G.~Grunberg,
\newblock Phys. Rev. D {\bf 29}, 2315 (1984).

\bibitem{Bjorken:1966jh}
J.~D. Bjorken,
\newblock Phys. Rev. {\bf 148}, 1467 (1966).

\bibitem{Bjorken:1969mm}
J.~D. Bjorken,
\newblock Phys. Rev. D {\bf 1}, 1376 (1970).

\bibitem{Aidala:2012mv}
C.~A. Aidala, S.~D. Bass, D.~Hasch and G.~K. Mallot,
\newblock Rev. Mod. Phys. {\bf 85}, 655 (2013).

\bibitem{Deur:2005cf}
A.~Deur, V.~Burkert, J.-P. Chen and W.~Korsch,
\newblock Phys. Lett. B {\bf 650}, 244 (2007).

\bibitem{Deur:2008rf}
A.~Deur, V.~Burkert, J.~P. Chen and W.~Korsch,
\newblock Phys. Lett. B {\bf 665}, 349 (2008).

\bibitem{Deur:2014vea}
A.~Deur {\em et~al.},
\newblock Phys. Rev. D {\bf 90}, 012009 (2014).

\bibitem{Ackerstaff:1997ws}
K.~Ackerstaff {\em et~al.},
\newblock Phys. Lett. B {\bf 404}, 383 (1997).

\bibitem{Ackerstaff:1998ja}
K.~Ackerstaff {\em et~al.},
\newblock Phys. Lett. B {\bf 444}, 531 (1998).

\bibitem{Airapetian:1998wi}
A.~Airapetian {\em et~al.},
\newblock Phys. Lett. B {\bf 442}, 484 (1998).

\bibitem{Airapetian:2002rw}
A.~Airapetian {\em et~al.},
\newblock Phys. Rev. Lett. {\bf 90}, 092002 (2003).

\bibitem{Airapetian:2006vy}
A.~Airapetian {\em et~al.},
\newblock Phys. Rev. D {\bf 75}, 012007 (2007).

\bibitem{Kim:1998kia}
J.~H. Kim {\em et~al.},
\newblock Phys. Rev. Lett. {\bf 81}, 3595 (1998).

\bibitem{Alexakhin:2006oza}
V.~{\relax Yu}. Alexakhin {\em et~al.},
\newblock Phys. Lett. B {\bf 647}, 8 (2007).

\bibitem{Alekseev:2010hc}
M.~G. Alekseev {\em et~al.},
\newblock Phys. Lett. B {\bf 690}, 466 (2010).

\bibitem{Adolph:2015saz}
C.~Adolph {\em et~al.},
\newblock Phys. Lett. B {\bf 753}, 18 (2016).

\bibitem{Anthony:1993uf}
P.~L. Anthony {\em et~al.},
\newblock Phys. Rev. Lett. {\bf 71}, 959 (1993).

\bibitem{Abe:1994cp}
K.~Abe {\em et~al.},
\newblock Phys. Rev. Lett. {\bf 74}, 346 (1995).

\bibitem{Abe:1995mt}
K.~Abe {\em et~al.},
\newblock Phys. Rev. Lett. {\bf 75}, 25 (1995).

\bibitem{Abe:1995dc}
K.~Abe {\em et~al.},
\newblock Phys. Rev. Lett. {\bf 76}, 587 (1996).

\bibitem{Abe:1995rn}
K.~Abe {\em et~al.},
\newblock Phys. Lett. B {\bf 364}, 61 (1995).

\bibitem{Anthony:1996mw}
P.~L. Anthony {\em et~al.},
\newblock Phys. Rev. D {\bf 54}, 6620 (1996).

\bibitem{Abe:1997cx}
K.~Abe {\em et~al.},
\newblock Phys. Rev. Lett. {\bf 79}, 26 (1997).

\bibitem{Abe:1997qk}
K.~Abe {\em et~al.},
\newblock Phys. Lett. B {\bf 404}, 377 (1997).

\bibitem{Abe:1997dp}
K.~Abe {\em et~al.},
\newblock Phys. Lett. B {\bf 405}, 180 (1997).

\bibitem{Abe:1998wq}
K.~Abe {\em et~al.},
\newblock Phys. Rev. {\bf D58}, 112003 (1998).

\bibitem{Anthony:1999py}
P.~L. Anthony {\em et~al.},
\newblock Phys. Lett. {\bf B458}, 529 (1999).

\bibitem{Anthony:1999rm}
P.~L. Anthony {\em et~al.},
\newblock Phys. Lett. B {\bf 463}, 339 (1999).

\bibitem{Anthony:2000fn}
P.~L. Anthony {\em et~al.},
\newblock Phys. Lett. B {\bf 493}, 19 (2000).

\bibitem{Anthony:2002hy}
P.~L. Anthony {\em et~al.},
\newblock Phys. Lett. B {\bf 553}, 18 (2003).

\bibitem{Gribov:1972}
V.~N. Gribov and L.~N. Lipatov,
\newblock Sov. J. Nucl. Phys. {\bf 15}, 438 (1972).

\bibitem{Altarelli:1977}
G.~Altarelli and G.~Parisi,
\newblock Nucl. Phys. B {\bf 126}, 298 (1977).

\bibitem{Dokshitzer:1977}
Y.~L. Dokshitzer,
\newblock Sov. Phys. JETP {\bf 46}, 641 (1977).

\bibitem{Kataev:1994gd}
A.~L. Kataev,
\newblock Phys. Rev. D {\bf 50}, R5469 (1994).

\bibitem{Baikov:2010je}
P.~A. Baikov, K.~G. Chetyrkin and J.~H. Kuhn,
\newblock Phys. Rev. Lett. {\bf 104}, 132004 (2010).

\bibitem{Chang:2011ei}
L.~Chang and C.~D. Roberts,
\newblock Phys. Rev. C {\bf 85}, 052201(R) (2012).

\bibitem{Chang:2013pq}
L.~Chang {\em et~al.},
\newblock Phys. Rev. Lett. {\bf 110}, 132001 (2013).

\bibitem{Shi:2015esa}
C.~Shi {\em et~al.},
\newblock Phys. Rev. D {\bf 92}, 014035 (2015).

\bibitem{Ding:2015rkn}
M.~Ding, F.~Gao, L.~Chang, Y.-X. Liu and C.~D. Roberts,
\newblock Phys. Lett. B {\bf 753}, 330 (2016).

\bibitem{Raya:2015gva}
K.~Raya {\em et~al.},
\newblock Phys. Rev. D {\bf 93}, 074017 (2016).

\bibitem{Raya:2016yuj}
K.~Raya, M.~Ding, A.~Bashir, L.~Chang and C.~D. Roberts,
\newblock (arXiv:1610.06575 [nucl-th]),
\newblock {\emph{Partonic structure of neutral pseudoscalars via two photon
  transition form factors}}.

\end{thebibliography}

\end{document}